# Spectrum Sharing in 6G Space-Ground Integrated Networks: A Ground Protection Zone-Based Design

\* Bodong Shang

Eastern Institute for Advanced Study

Eastern Institute of Technology, Ningbo

Ningbo, China

\* Corresponding author: bdshang@eitech.edu.cn

Xiangyu Li

Eastern Institute for Advanced Study

Eastern Institute of Technology, Ningbo

Ningbo, China

xyli@eitech.edu.cn

Zheng Wang

Ningbo Institute of Digital Twin

Eastern Institute of Technology, Ningbo

Ningbo, China

zwang@idt.eitech.edu.cn

Junchao Ma

School of Electrical and Information Engineering

Jiangsu University of Technology

Changzhou, China

junchao_ma@jstu.edu.cn

*Abstract*—Space-ground integrated network (SGIN) has been envisioned as a competitive solution for large scale and wide coverage of future wireless networks. By integrating both the non-terrestrial network (NTN) and the terrestrial network (TN), SGIN can provide high speed and omnipresent wireless network access for the users using the predefined licensed spectrums. Considering the scarcity of the spectrum resource and the low spectrum efficiency of the SGIN, we enable the NTN and TN to share the spectrum to improve overall system performance, i.e., weighted-sum area data rate (WS-ADR). However, mutual interference between NTN and TN is often inevitable and thus causes SGIN performance degradation. In this work, we consider a ground protection zone for the TN base stations, in which the NTN users are only allowed to use the NTN reserved spectrum to mitigate the NTN and TN mutual interference. We analytically derive the coverage probability and area data rate (ADR) of the typical users and study the performance under various protection zone sizes and spectrum allocation parameter settings. Simulation and numerical results demonstrate that the WS-ADR could be maximized by selecting the appropriate radius of protection zone and bandwidth allocation factor in the SGIN.

*Keywords-SGIN; Space-ground integrated networks; spectrum sharing; coverage probability; weighted-sum area data rate*

## I. INTRODUCTION

Future communication networks are envisioned to provide wireless services with high data rate, low latency and high reliability over a wide span of geographical areas [1]. In current terrestrial networks (TNs), the evolving 5th-generation (5G) can offer high-capacity data pipes to meet the above service demands [2]. However, the technologies inspired by the roadmap from 5G to 6G in the coming decade are also calling for high efficiency, high flexibility for future networks with optimal coverage. Since the TN infrastructures are fixed after construction, they can hardly bring the expected performance leap from 5G to 6G without the assistance of other network components [3][4]. Thus, the improvement of conventional TN is substantial in future wireless technologies.

Recently, space-ground integrated network (SGIN) has attracted intensive attention in related research works, which is deemed as a key component in 6G communication networks [5]. By incorporating the merits of non-terrestrial satellites into the existing TN, SGIN can provide a large-scale seamless coverage of geographical areas, which is beyond the capabilities of the TN. Additionally, since the satellites are non-static in space, the mobility of the non-terrestrial network (NTN) naturally leads to dynamic network topologies and thus a high flexibility of the SGIN. Meanwhile, with the NTN, the self-organization of the SGIN can be implemented and the TN infrastructures can intelligently collaborate with each other and concentrate on their dominant services [6].

However, SGIN also has its own challenges when the NTN is incorporated to enhance the TN performance. On the one hand, due to the randomness of the non-terrestrial satellite locations, the SGIN topology is usually highly dynamic with a significant impact on the network performance, including the coverage, the data rate as well as the network reliability [7]. Hence, with high uncertainties in the SGIN architectures and topologies, precise modeling of the network is essential to the SGIN performance evaluation process. On the other hand, the spectrum in SGIN is a scarce resource. Since additional non-terrestrial nodes are incorporated in the system, they often need to share the limited spectrum with the TN nodes such as the base stations (BSs), and mutual interference is thus inevitable [8]. This usually incurs severe system performance degradation and eventually leads to device outage when the interference is not tolerable. Thus, a dedicated protection mechanism for the network devices emerges as the second major challenge.

Regarding spectrum sharing in wireless networks, the concept of the protection zone has been introduced to isolate the secondary users from the primary users and prevent harmful interference. The protection zone is normally a sphere centered at the BS with a certain radius. While an NTN user resides outside any of the protection zones in the geographical area of interest, it can access the spectrum shared by the TN and the NTN, as the interference it receives is small enough and neglectable. Otherwise, it needs to access the spectrum reserved for the NTN to avoid interference. The protection zone size and the spectrum allocation for both TN and NTN have significant impacts on the SGIN performance, and guidance on the related parameter selection is necessary. To the best of our knowledge, none of the previous work has addressed the issues above. Hence in this paper, we aim to model the SGIN above and give a detailed study of network performance and discuss the parameter selection issues.

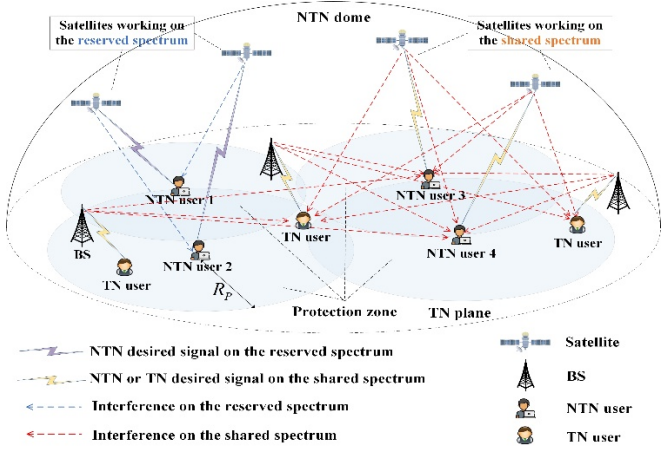

Fig. 1. Ground protection zone-based spectrum sharing in SGIN

The rest of the paper is organized as follows. In Section II, we introduce the SGIN system model with TN/NTN shared spectrum and NTN reserved spectrum and formulate the objective. In Section III, we give detailed theoretical analysis of the area data rates for the users in both shared spectrum and reserved spectrum. In Section IV, we conduct simulation and numerical experiments to study the coverage probability of typical users and their weighted-sum area data rate (WS-ADR). Finally, the concluding remarks are given in Section V.

## II. SYSTEM MODEL AND OBJECTIVE

In this section, we introduce the system model and objective of the considered protection zone-based spectrum sharing in the SGIN.

### A. Network Model

As shown in Fig. 1, there are multiple satellites and terrestrial BSs. Without loss of generality, we use a homogeneous spherical Poisson point process (SPPP) to model the distribution of satellites at a certain altitude $H$ with a density of $\lambda_N$. The set of satellites is denoted by $\Phi_N$. Considering that the signal attenuates significantly in TN due to non-line-of-sight (NLoS) over several kilometers, we consider a limited region of interest for TN, and thus we assume that BSs are distributed on a plane following another independent homogeneous Poisson point process (PPP) with the density of $\lambda_T$ and the set of BSs is denoted by $\Phi_T$. Ground users are classified into two independent sets for NTN users and TN users modeled by two independent PPPs with the densities of $\lambda_{NU}$ and $\lambda_{TU}$, respectively. The assumptions of random distribution will be validated by simulations under real satellite constellations. It's worth noting that even though the network elements (e.g., satellites, BSs) are distributed on a three-dimensional sphere, we use these sets to represent only the network elements within the visual range of a typical user.

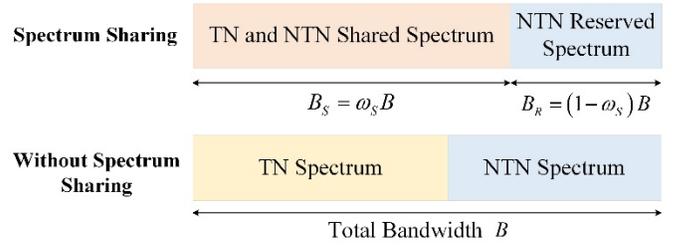

Fig. 2. An illustration of spectrum sharing model

### B. Spectrum Sharing Model

As shown in Fig. 2, in the proposed spectrum sharing policy, the total bandwidth is divided into two parts, namely shared spectrum and reserved spectrum. The bandwidth of shared spectrum is $B_S = \omega_S B$, where $\omega_S \in [0,1]$ is the factor indicating the fraction of total bandwidth $B$ could be used for users in the shared spectrum. The bandwidth of reserved spectrum is $B_R = (1-\omega_S)B$.

We consider that NTN users are classified into two sets, i.e., the set of NTN users using the shared spectrum denoted by $\Phi_{NU,S}$ and the set of NTN users using the reserved spectrum denoted by $\Phi_{NU,R}$. If no BS is located within the protection distance $R_P$ away from a ground NTN user, the NTN user can use the shared spectrum. Otherwise, the NTN user uses the reserved spectrum. By adjusting $R_P$, we can control the interference generated from NTN to TN, and thus the system overall performance. The set of TN users is $\Phi_{TU}$. Therefore, users in $\Phi_{NU,S}$ and $\Phi_{TU}$ utilize the shared spectrum, while users in $\Phi_{NU,R}$ enjoy the reserved spectrum.

Accordingly, the set of satellites with their associated users in $\Phi_{NU,S}$ is denoted by $\Phi_{N,S}$. The set of satellites with their associated users in $\Phi_{NU,R}$ is denoted by $\Phi_{N,R}$.

In Fig. 2, for the scenario without spectrum sharing, TN and NTN use their dedicated spectrum, and there is no interference between NTN and TN. However, the bandwidth for each network is limited and, in some places, there would be a waste of spectrum resources, which motivates us to consider a more flexible and efficient spectrum utilization policy.

### C. Channel Model

The wireless transmission in either NTN or TN is characterized by a composite model encompassing distance dependent path-loss and small-scale fading. We assume that each satellite or BS serves one user in a time slot. For a typical NTN user $u_{NU,S}$ in $\Phi_{NU,S}$ located at $(0,0,R_E)$, where $R_E$ is the radius of the earth, the received signal of $u_{NU,S}$ is given by

$$y_{NU,S} = \sum_{i \in \Phi_{N,S}} \sqrt{P_N G_{\text{ml/sl}} \beta_{N,0} r_i^{-\alpha_N}} h_i s_i \\ + \sum_{j \in \Phi_T} \sqrt{P_T G_T \beta_{T,0} l_j^{-\alpha_T}} g_j b_j + n_0, \quad (1)$$

where in the first term, $P_N$ is the satellite's transmission power, $G_{\text{ml/sl}}$ is satellite's antenna gain for main lobe (ml) or side lobe (sl) beam, $\beta_{N,0}$ represents the reference path-loss, $r_i$ is the distance between $u_{NU,S}$ and the $i$-th satellite (we use the number $i$ to express the ordinal number and the entity interchangeably), $\alpha_N$ is the path-loss exponent in NTN, $h_i \sim \text{Nakagami}(m, \Omega)$ is the small-scale fading with a shape and scale parameter, i.e., $m$ and $\Omega$, respectively, $s_i$ is the transmitted symbol at the $i$-th satellite and $\mathbb{E}\{|s_i|^2\} = 1$, and $\mathbb{E}\{\cdot\}$ denotes the expectation calculation. In the second term of (1), $P_T$ is the BS's transmission power, $G_T$ is BS's antenna gain, $\beta_{T,0}$ represents the reference path-loss in TN, $l_{NU,j}$ is the distance between $u_{NU,S}$ and the $j$-th BS, $\alpha_T$ is the path-loss exponent in TN, $g_i$ is the small-scale fading following a Rayleigh distribution, $b_j$ is the transmitted symbol at the $j$-th BS and $\mathbb{E}\{|b_j|^2\} = 1$. $n_0$ is the additive white Gaussian noise with a variance of $\sigma_0^2$. The SINR of the $i$-th $u_{NU,S}$ is given by

$$\text{SINR}_{NU,S}^i = \frac{P_N G_{\text{ml}} \beta_{N,0} r_i^{-\alpha_N} |h_i|^2}{\sum\limits_{\substack{i' \in \Phi_{N,S} \\ i' \neq i}} P_N G_{\text{sl}} \beta_{N,0} r_{i'}^{-\alpha_N} |h_{i'}|^2 + \sum\limits_{j \in \Phi_T} P_T G_T \beta_{T,0} l_{NU,j}^{-\alpha_T} |g_j|^2 + \sigma_0^2}. \quad (2)$$

For a typical NTN user $u_{NU,R}$ in $\Phi_{NU,R}$ using the reserved spectrum $B_R$, the received signal of $u_{NU,R}$ is given by

$$y_{NU,R} = \sum_{i \in \Phi_{N,R}} \sqrt{P_N G_{\text{ml/sl}} \beta_{N,0} r_i^{-\alpha_N}} h_i s_i + n_0, \quad (3)$$

where the signals are generated from $\Phi_{NU,R}$, and there is no interference signal from TN. The SINR of the $i$-th $u_{NU,R}$ is

$$\text{SINR}_{NU,R}^i = \frac{P_N G_{\text{ml}} \beta_{N,0} r_i^{-\alpha_N} |h_i|^2}{\sum\limits_{i' \in \Phi_{N,R}, i' \neq i} P_N G_{\text{sl}} \beta_{N,0} r_{i'}^{-\alpha_N} |h_{i'}|^2 + \sigma_0^2}. \quad (4)$$

For a typical user $u_{TU}$ in $\Phi_{TU}$ using the reserved spectrum $B_R$, the received signal of $u_{TU}$, i.e., $y_{TU}$, is similar to $y_{NU,S}$. The SINR of the $j$-th $u_{TU}$ is given by

$$\text{SINR}_{TU}^j = \frac{P_T G_T \beta_{T,0} l_{TU,j}^{-\alpha_T} |g_j|^2}{\sum\limits_{i \in \Phi_{N,S}} P_N G_{\text{sl}} \beta_{N,0} r_i^{-\alpha_N} |h_i|^2 + \sum\limits_{\substack{j' \in \Phi_T \\ j' \neq j}} P_T G_T \beta_{T,0} l_{TU,j'}^{-\alpha_T} |g_{j'}|^2 + \sigma_0^2}, \quad (5)$$

where $l_{TU,j}$ is the distance between $u_{TU}$ and the $j$-th BS.

It is worth noting that there are three types of typical users in this paper, i.e., the users in $\Phi_{NU,S}$, $\Phi_{NU,R}$, and $\Phi_{TU}$.

### D. Performance Metric

Since we divide total spectrum into shared spectrum and reserved spectrum, the amount of spectrum should be counted in each type of user's performance metric. Therefore, we propose to consider the area data rate (ADR) for each type of user. Specifically, the ADR of a typical user in $\Phi_{NU,S}$ is

$$\text{ADR}_{N,S} = \lambda_{N,S} B_S \mathbb{E}\{\log_2(1 + \text{SINR}_{NU,S}^i)\}, \quad (6)$$

where $\lambda_{N,S}$ is the density of satellites serving users in $\Phi_{NU,S}$. Accordingly, the ADR of a typical user in $\Phi_{NU,R}$ is given by

$$\text{ADR}_{N,R} = \lambda_{N,R} B_R \mathbb{E}\{\log_2(1 + \text{SINR}_{NU,R}^i)\}, \quad (7)$$

where $\lambda_{N,R}$ is the density of satellites serving users in $\Phi_{NU,R}$. The ADR of a typical user in $\Phi_{TU}$ is given by

$$\text{ADR}_T = \frac{\lambda_T}{\delta} B_S \mathbb{E}\{\log_2(1 + \text{SINR}_{TU}^j)\}, \quad (8)$$

where $\delta$ is the frequency reuse factor.

Now, we are in the position of defining overall system metric, i.e., weighted-sum ADR (WS-ADR), as follows

$$\text{WS-ADR} = \xi_{N,S} \text{ADR}_{N,S} + \xi_{N,R} \text{ADR}_{N,R} + \xi_T \text{ADR}_T, \quad (9)$$

where $\xi_{N,S}$, $\xi_{N,R}$, and $\xi_T$ denote the weights.

### E. Objective

Our objective is to maximize the WS-ADR by selecting appropriate radius of ground protection zone $R_P$ and bandwidth allocation factor $\omega_S$ under various network conditions while guaranteeing each type of user's quality of service (QoS), as follows

$$\begin{aligned} & \max_{R_P, \omega_S} \text{ WS-ADR} \\ & \text{s.t. } \text{ADR}_{N,S} \geq \text{ADR}_{N,S}^{th} \\ & \quad \text{ADR}_{N,R} \geq \text{ADR}_{N,R}^{th} \\ & \quad \text{ADR}_T \geq \text{ADR}_T^{th} \end{aligned} \quad (10)$$

where $\text{ADR}_{N,S}^{th}$, $\text{ADR}_{N,R}^{th}$, and $\text{ADR}_T^{th}$ are the ADR thresholds of users in $\Phi_{NU,S}$, $\Phi_{NU,R}$, and $\Phi_{TU}$, respectively.

## III. ANALYSIS OF SPECTRUM SHARING IN SGIN

In this section, we derive the coverage probabilities and the ADRs of three types of typical users in $\Phi_{NU,S}$, $\Phi_{NU,R}$, and $\Phi_{TU}$, respectively. The results will be used for calculating the WS-ADR in (9).

### A. ADR of NTN Users in the Shared Spectrum

The coverage probability of the $i$-th typical user $u_{NU,S}$ is given by

$$\Pr\{\text{SINR}_{NU,S}^i \geq \gamma_{NU,th}\} = \Pr\left\{|h_i|^2 \geq \frac{\gamma_{NU,th}\left(I_{N-N,S}^i + I_{T-N} + \sigma_0^2\right)}{P_N G_{ml}\beta_{N,0} r_i^{-\alpha_N}}\right\}, \quad (11)$$

where $\gamma_{NU,th}$ is the SINR threshold of a NTN user, $I_{N-N,S}^i$ and $I_{T-N}$ indicate the interference powers generated from NTN to NTN on the shared spectrum and generated from TN to NTN on the shared spectrum, respectively. The detailed expressions are given as follows

$$I_{N-N,S}^i = \sum_{i' \in \Phi_{N,S}, i' \neq i} P_N G_{sl}\beta_{N,0} r_{i'}^{-\alpha_N} |h_{i'}|^2,$$
$$I_{T-N} = \sum_{j \in \Phi_T} P_T G_T \beta_{T,0} d_j^{-\alpha_T} |g_j|^2. \quad (12)$$

Since $h_i$ follows a Nakagami-$m$ distribution, $|h_i|^2$ follows a Gamma distribution, and we have the following results

$$\Pr\{\text{SINR}_{NU,S}^i \geq \gamma_{NU,th}\}$$
$$= \Pr\left\{|h_i|^2 \geq \frac{\gamma_{NU,th}\left(I_{N-N,S}^i + I_{T-N} + \sigma_0^2\right)}{P_N G_{ml}\beta_{N,0} r_i^{-\alpha_N}}\right\}$$
$$\stackrel{(a)}{=} \mathbb{E}\left\{e^{-s\left(I_{N-N,S}^i + I_{T-N} + \sigma_0^2\right)} \sum_{k=0}^{m-1} \frac{\left[s\left(I_{N-N,S}^i + I_{T-N} + \sigma_0^2\right)\right]^k}{k!}\right\}$$
$$= \mathbb{E}\left\{\sum_{k=0}^{m-1} \frac{s^k}{k!}\left(I_{N-N,S}^i + I_{T-N} + \sigma_0^2\right)^k e^{-s\left(I_{N-N,S}^i + I_{T-N} + \sigma_0^2\right)}\right\}$$
$$= \int_{R_{min}}^{R_{max}} \sum_{k=0}^{m-1} \frac{s^k}{k!}(-1)^k \left.\frac{\partial^k \mathcal{L}_{I_{N-N,S}^i + I_{T-N} + \sigma_0^2}(s)}{\partial s^k}\right|_{s=m\frac{\gamma_{NU,th} r^{\alpha_N}}{P_N G_{ml}\beta_{N,0}}} f_R(r) dr, \quad (13)$$

where in (a), we have $s = m\frac{\gamma_{NU,th} r^{\alpha_N}}{P_N G_{ml}\beta_{N,0}}$ for notation simplicity, and we utilize complementary cumulative distribution function (CCDF) of channel power gain $|h_i|^2$ in the NTN as follows

$$\Pr\{|h_i|^2 \geq x\} = e^{-mx}\sum_{k=0}^{m-1} \frac{(mx)^k}{k!} \quad (14)$$

where we suppose $\Omega = \mathbb{E}\{|h_i|^2\} = 1$.

In the last step of (13), $\mathcal{L}_{I_{N-N,S}^i + I_{T-N} + \sigma_0^2}$ denotes the Laplace transform of the term $I_{N-N,S}^i + I_{T-N} + \sigma_0^2$. We have the following property

$$\mathcal{L}_{I_{N-N,S}^i + I_{T-N} + \sigma_0^2} = \mathcal{L}_{I_{N-N,S}^i} \mathcal{L}_{I_{T-N}} e^{-s\sigma_0^2}. \quad (15)$$

The Laplace transform of $I_{N-N,S}^i$, i.e., $\mathcal{L}_{I_{N-N,S}^i}$, is given by

$$\mathcal{L}_{I_{N-N,S}^i}(s) = \mathbb{E}\left\{e^{-sI_{N-N,S}^i}\right\}$$
$$= \mathbb{E}\left\{\exp\left(-s\sum_{i' \in \Phi_{N,S}, i' \neq i} P_N G_{sl}\beta_{N,0} r_{i'}^{-\alpha_N}|h_{i'}|^2\right)\right\}$$
$$\stackrel{(a)}{=} \exp\left(-\lambda_{N,S}\int_{r \in \mathcal{A}_r}\left(1 - \mathbb{E}\left\{e^{-sP_N G_{sl}\beta_{N,0}|h_i|^2 r^{-\alpha_N}}\right\}\right) dr\right)$$
$$\stackrel{(b)}{=} \exp\left(-\lambda_{N,S}\int_{r \in \mathcal{A}_r}\left[1 - \frac{1}{\left(1 + P_N G_{sl}\beta_{N,0}\frac{sr^{-\alpha_N}}{m}\right)^m}\right] dr\right)$$
$$\stackrel{(c)}{=} \exp\left(-2\pi\lambda_{N,S}\frac{R_S}{R_E}\int_r^{R_{max}}\left[1 - \frac{1}{\left(1 + P_N G_{sl}\beta_{N,0}\frac{sr^{-\alpha_N}}{m}\right)^m}\right] r dr\right)$$
$$\stackrel{(d)}{=} \exp\left(-\lambda_{N,S}\pi\frac{R_S}{R_E}\left(P_N G_{sl}\beta_{N,0}\frac{s}{m}\right)^{\frac{2}{\alpha_N}}\right.$$
$$\left.\cdot \int_{\left(P_N G_{sl}\beta_{N,0}\frac{s}{m}\right)^{-\frac{2}{\alpha_N}} R_{max}^2}^{\left(P_N G_{sl}\beta_{N,0}\frac{s}{m}\right)^{-\frac{2}{\alpha_N}} r^2}\left[1 - \frac{1}{\left(1 + u^{-\frac{\alpha_N}{2}}\right)^m}\right] du\right), \quad (16)$$

where in (a) we use the probability generating functional (PGFL) of PPP, (b) follows from the Gamma distribution of the NTN channel power gain and $\mathcal{A}_r$ is the area on the spherical cap within the distance to the typical NTN user between $r$ and $R_{max} = \sqrt{R_S^2 - R_E^2}$, (c) is from the derivative, i.e., $\partial|\mathcal{A}_r|/\partial r = 2\pi r R_S/R_E$, (d) is from the change of variable, i.e., $u = \left(P_N G_{sl}\beta_{N,0}\frac{s}{m}\right)^{-2/\alpha_N} r^2$. Since NTN users are randomly distributed on the earth surface, the distribution of satellites on the shared spectrum could be approximated by a SPPP with the density of $\lambda_{N,S}$ given as follows

$$\lambda_{N,S} = \Pr_{NU,S}\frac{\lambda_N}{\delta} = \exp(-\pi\lambda_T R_P^2)\frac{\lambda_N}{\delta}, \quad (17)$$

where $\Pr_{NU,S} = \exp(-\pi\lambda_T R_P^2)$ denotes the probability that no BS exists in the NTN user's protection zone.

The Laplace transform of interference power generated from TN to NTN on the shared spectrum, i.e., $\mathcal{L}_{I_{T-N}}$, is given by

$$\mathcal{L}_{I_{T-N}}(s) = \mathbb{E}\{e^{-sI_{T-N}}\}$$
$$= \mathbb{E}\left\{\exp\left(-s\sum_{j\in\Phi_T} P_T G_T \beta_{T,0} l_{NU,j}^{-\alpha_T}|g_j|^2\right)\right\} \quad (18)$$
$$\overset{(a)}{=} \exp\left(-2\pi\frac{\lambda_T}{\delta}\int_{R_P}^{\infty}\left(1-\frac{1}{1+P_T G_T \beta_{T,0}sx^{-\alpha_T}}\right)xdx\right),$$

where in (a) we consider the interfering BSs are located at least $R_P$ distance away from the typical NTN user $u_{NU,S}$.

We assume that each NTN user is associated with its nearest satellite. The probability density function (PDF) of the desired link distance $r$ is given by

$$f_{R|\Phi_N(\mathcal{A})>1}(r) = 2\pi\lambda_N \frac{R_S}{R_E} \frac{e^{\lambda_N \pi \frac{R_S}{R_E}(R_S^2 - R_E^2)}}{e^{2\lambda_N \pi R_S(R_S - R_E)} - 1} r e^{-\lambda_N \pi \frac{R_S}{R_E}r^2}, \quad (19)$$

where $R_{\min} \leq r \leq R_{\max}$, $R_{\min} = R_S - R_E$.

By combining (15), (16), (17), (18), (19) into (13), we obtain the coverage probability of the typical user $u_{NU,S}$. The ADR of the NTN users on shared spectrum, i.e., (6), is calculated based on (13), as follows

$$\text{ADR}_{N,S} = \lambda_{N,S} B_S \int_0^{\infty} \Pr\{\text{SINR}_{NU,S}^i \geq 2^t - 1\}dt. \quad (20)$$

Note that $\text{ADR}_{N,S}$ in (20) is a function of $R_P$ and $\omega_S$.

### B. ADR of NTN Users in the Reserved Spectrum

The coverage probability of the $i$-th typical user $u_{NU,R}$ is given by

$$\Pr\{\text{SINR}_{NU,R}^i \geq \gamma_{NU,th}\}$$
$$= \Pr\left\{|h_i|^2 \geq \frac{\gamma_{NU,th}(I_{N-N,R}^i + \sigma_0^2)}{P_N G_{ml}\beta_{N,0}r_i^{-\alpha_N}}\right\} \quad (21)$$
$$= \int_{R_{\min}}^{R_{\max}} \sum_{k=0}^{m-1} \frac{s^k}{k!}(-1)^k \left.\frac{\partial^k \mathcal{L}_{I_{N-N,R}^i + \sigma_0^2}(s)}{\partial s^k}\right|_{s=m\frac{\gamma_{NU,th}r^{\alpha_N}}{P_N G_{ml}\beta_{N,0}}} f_R(r)dr.$$

Similar to (15) and (16), $\mathcal{L}_{I_{N-N,R}^i + \sigma_0^2}(s)$ is given by

$$\mathcal{L}_{I_{N-N,R}^i + \sigma_0^2}(s) = \exp\left(\begin{array}{c} -s\sigma_0^2 - \lambda_{N,R}\pi\frac{R_S}{R_E}\left(P_N G_{sl}\beta_{N,0}\frac{s}{m}\right)^{\frac{2}{\alpha_N}} \\ \cdot \int_{(P_N G_{sl}\beta_{N,0}\frac{s}{m})^{-\frac{2}{\alpha_N}}r^2}^{(P_N G_{sl}\beta_{N,0}\frac{s}{m})^{-\frac{2}{\alpha_N}}R_{\max}^2} 1-\frac{1}{\left(1+u^{-\frac{\alpha_N}{2}}\right)^m}du \end{array}\right), \quad (22)$$

where the density of satellites on the reserved spectrum is

$$\lambda_{N,R} = \frac{\lambda_N}{\delta} - \lambda_{N,S} = [1-\exp(-\pi\lambda_T R_P^2)]\frac{\lambda_N}{\delta}. \quad (23)$$

The ADR of the NTN users on reserved spectrum, i.e., (7), is calculated based on (21), as follows

$$\text{ADR}_{N,R} = \lambda_{N,R} B_R \int_0^{\infty} \Pr\{\text{SINR}_{NU,R}^i \geq 2^t - 1\}dt. \quad (24)$$

### C. ADR of TN Users in the Shared Spectrum

The coverage probability of the $j$-th typical user $u_{TU}$ is

$$\Pr\{\text{SINR}_{TU}^j \geq \gamma_{TU,th}\}$$
$$= \Pr\left\{|g_j|^2 \geq \frac{\gamma_{TU,th}(I_{N-T,S} + I_{T-T}^j + \sigma_0^2)}{P_T G_T \beta_{T,0} l_{TU,j}^{-\alpha_T}}\right\} \quad (25)$$
$$= \mathbb{E}\left\{\exp\left(-\frac{\gamma_{TU,th}l_{TU,j}^{\alpha_T}}{P_T G_T \beta_{T,0}}(I_{N-T,S} + I_{T-T}^j + \sigma_0^2)\right)\right\}$$
$$= \int_0^{\infty} \mathcal{L}_{I_{N-T,S}}(s)\mathcal{L}_{I_{T-T}^j}(s)e^{-s\sigma_0^2}\bigg|_{s=\frac{\gamma_{TU,th}l_{TU,j}^{\alpha_T}}{P_T G_T \beta_{T,0}}} f_L(l)dl,$$

where the notations of interference are expressed as

$$I_{N-T,S} = \sum_{i\in\Phi_{N,S}} P_N G_{sl}\beta_{N,0}r_i^{-\alpha_N}|h_i|^2,$$
$$I_{T-T}^j = \sum_{\substack{j'\in\Phi_T \\ j'\neq j}} P_T G_T \beta_{T,0} l_{TU,j'}^{-\alpha_T}|g_{j'}|^2. \quad (26)$$

Moreover, the Laplace transform $\mathcal{L}_{I_{N-T,S}}(s)$ is given by

$$\mathcal{L}_{I_{N-T,S}}(s) \approx \exp\left(\begin{array}{c} -\lambda_{N,S}\pi\frac{R_S}{R_E}\left(P_N G_{sl}\beta_{N,0}\frac{s}{m}\right)^{\frac{2}{\alpha_N}} \\ \cdot \int_{(P_N G_{sl}\beta_{N,0}\frac{s}{m})^{-\frac{2}{\alpha_N}}R_{\min}^2}^{(P_N G_{sl}\beta_{N,0}\frac{s}{m})^{-\frac{2}{\alpha_N}}R_{\max}^2} 1-\frac{1}{\left(1+u^{-\frac{\alpha_N}{2}}\right)^m}du \end{array}\right), \quad (27)$$

and the Laplace transform $\mathcal{L}_{I_{T-T}^j}(s)$ is given by

$$\mathcal{L}_{I_{T-T}^j}(s) = \mathbb{E}\{e^{-sI_{T-T}^j}\}$$
$$= \exp\left(-2\pi\frac{\lambda_T}{\delta}\int_{l_{TU,j}}^{\infty}\left(1-\frac{1}{1+P_T G_T \beta_{T,0}sx^{-\alpha_T}}\right)xdx\right), \quad (28)$$

where $\mathcal{L}_{I_{T-T}^j}(s)$ is a function of $l_{TU,j}$ whose PDF is given by

$$f_L(l) = 2\pi\lambda_T l e^{-\pi\lambda_T l^2}, \ 0 \leq l \leq \infty. \quad (29)$$

Thus, by combining (27), (28), (29) into (25), we obtain the coverage probability of $u_{TU}$. The ADR of the TN users is

$$\text{ADR}_T = \frac{\lambda_T}{\delta} B_S \int_0^{\infty} \Pr\{\text{SINR}_{TU}^j \geq 2^t - 1\}dt. \quad (30)$$

By combining (20), (24), (30) into (9), we get the WS-ADR.

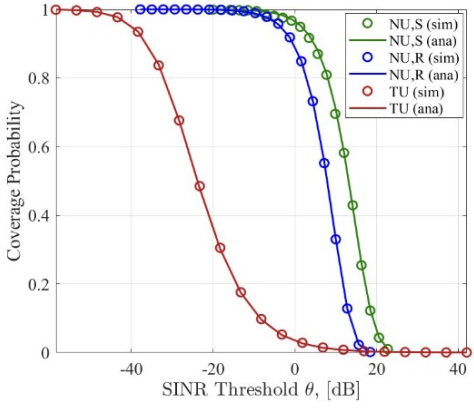

Fig. 3. Coverage probability of typical users.

## IV. SIMULATION AND DISCUSSIONS

In this section, we assume there is a $100 \text{ km} \times 100 \text{ km}$ remote area where BSs are distributed with a density of $\lambda_T = 0.318/\text{km}^2$. The transmit power, antenna gain, service radius are $P_T = 46 \text{ dBm}$, $G_T = 20 \text{ dBi}$, and $R_T = 10 \text{ km}$, respectively. The reference path-loss at 1 m is $\beta_{0,T} = 1$ and the path-loss exponent is $\alpha_T = 3.5$. In the NTN, the satellites are distributed at altitude $H = 500 \text{ km}$ with a density of $\lambda_N = 1 \times 10^{-5}/\text{km}^2$. The transmit power, main-lobe gain and side-lobe gain are $P_N = 50 \text{ dBm}$, $G_{ml} = 30 \text{ dBi}$, $G_{sl} = 20 \text{ dBi}$, respectively. The reference path-loss at 1 m is $\beta_{0,N} = 1$ and the path-loss exponent is $\alpha_N \approx 2.0$. We denote $m = 1$, $\Omega = 1$ for the Nakagami-$m$ channel, and the frequency reuse factor $\delta = 2$. The TN bandwidth $B_{TN} = 200 \text{ MHz}$, the NTN bandwidth $B_{NTN} = 100 \text{ MHz}$, and the total bandwidth $B = 300 \text{ MHz}$.

In Fig. 3, coverage probabilities of the typical user $u_{NU,S}$, $u_{NU,R}$, and $u_{TU}$ are plotted. The radius of protection zone is set as $R_P = 12 \text{ km}$. The analysis results match closely with simulation. Specifically, in the given range of SINR thresholds, $u_{NU,S}$ has the highest coverage, and the coverage of both $u_{NU,S}$ and $u_{NU,R}$ is significantly higher than that of $u_{TU}$. This is because less satellites are engaged in the shared spectrum than those in reserved spectrum. The use of protection zone effectively reduces the interference from the nearest BS in TNs.

The ADR for each of the three users, and their WS-ADR are presented in Fig. 4. Each user has its own optimal ADR, e.g., $u_{NU,S}$ obtains its optimal ADR when $\omega_S$ is large and $R_P$ is small, $u_{NU,R}$ achieves its highest ADR with a smaller $\omega_S$ but a larger $R_P$, and $u_{TU}$ gets its maximum ADR with both larger $\omega_S$ and larger $R_P$. Their weights, however, are determined by engineering requirements, and herein we have $\xi_{N,S}$, $\xi_{N,R}$, and $\xi_T$ set as 27, 29, and 0.9, respectively. It is found that there exist multiple combinations of $R_P$ and $\omega_S$ to obtain highest WS-ADR, which exhibits a trade-off.

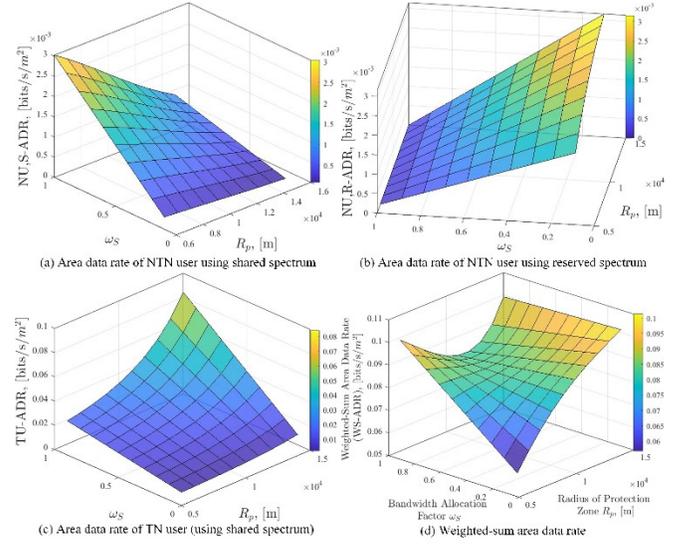

(a) Area data rate of NTN user using shared spectrum
(b) Area data rate of NTN user using reserved spectrum
(c) Area data rate of TN user (using shared spectrum)
(d) Weighted-sum area data rate

Fig. 4. ADR of typical users and their WS-ADR.

## V. CONCLUSIONS

In this paper, we investigated the SGIN model with both the shared spectrum and the NTN reserved spectrum, formulating the ADR of typical users based on their protection zone sizes. We provided analytical expressions of coverage probability and ADR for each typical user and the WS-ADR of the network. There exists a trade-off between bandwidth allocation factor and radius of protection zone to achieve an optimal WS-ADR.